\documentstyle[12pt,preprint,aps,epsf,floats,amssymb]{revtex}
\def\NPB#1#2#3{Nucl. Phys. B\,{\bf #1}, #3 (19#2)}

\def\PRD#1#2#3{Phys. Rev. D\,{\bf #1}, #3 (19#2)}
\def\PRDMM#1#2#3{Phys. Rev. D\,{\bf #1}, #3 (20#2)}
\def\PRL#1#2#3{Phys. Rev. Lett. {\bf#1}, #3 (19#2)}
\def\PRLMM#1#2#3{Phys. Rev. Lett. {\bf#1}, #3 (20#2)}

\def\EPJCMM#1#2#3{{Eur. Phys. J. direct C\,} {\bf #1}, #3 (20#2)}

\def\[{\left[}
\def\]{\right]}
\def\({\left(}
\def\){\right)}

\def\U1em{{U(1)_{\rm em}}}

\def\95CL{{95\%\,{\rm C.L.}}}

\def\End{\end{document}}

\newcommand{\PSbox}[3]{\mbox{\rule{0in}{#3}
\includegraphics{#1}\hspace{#2}}}

\tighten
\begin{document}
\preprint{
\noindent
\begin{minipage}[t]{3in}
\begin{flushleft}
\end{flushleft}
\end{minipage}
\hfill
\begin{minipage}[t]{3in}
\begin{flushright} 
CALT-68-2326\\
hep-ph/0104169\\
\vspace*{.7in}
\end{flushright}
\end{minipage}
}

\draft

\title{
Supersymmetric Correction to Top Quark Pair Production near Threshold}

\author{{\sc Shufang Su}\footnote{shufang@theory.caltech.edu} and~~ 
        {\sc Mark B. Wise}\footnote{wise@theory.caltech.edu}  
\vspace*{.2in}
}
\address{California Institute of Technology, Pasadena, California 91125
\vspace*{.2in}}


\maketitle

\begin{abstract}
We studied the leading supersymmetric  contribution to  top-antitop 
threshold production using the NRQCD framework.  
The one-loop matching to the 
potential and the Wilson coefficient of the 
leading $^3S_1$ production current were considered.
We point out that the leading correction to the 
potential is zero due to ${\rm SU}(3)_c$ gauge invariance.
This is true in general for any new physics that
enters above the electroweak scale.  The shape of the top quark pair 
production cross section is therefore almost unaffected near threshold, 
allowing
a precise determination of the top quark mass based on the 
Standard Model calculations.  
The supersymmetric correction to the Wilson 
coefficient $c_1$ of the production current decouples for heavy 
super particles.  Its contribution is smaller than the Standard Model 
next-to-next-leading-log results. 

\end{abstract}
\pacs{PACS numbers:14.65.Ha, 12.60.Jv, 12.38.Bx.}

It is likely that top quark pair production near threshold
will be investigated intensively in the future 
at an $e^+e^-$ linear collider.  
It can be used to determine the 
top quark mass, decay width, top Yukawa coupling and strong 
coupling constant.  The Standard Model (SM) QCD prediction has been 
studied in great detail by several groups \cite{top,iain1} in the framework of 
NRQCD (nonrelativistic QCD) \cite{nrqcd} up to next-to-next-leading-log (NNLL)
order.  Near the 
threshold, the velocity of the top quark
$v\sim\alpha_s\sim{0.15}$ is small.
However, the relevant energy scales
$m_t\sim{175}$ GeV, $m_tv\sim{25}$ GeV, $m_tv^2\sim{4}$GeV, together 
with top decay width $\Gamma_t\approx{1.5}$ GeV are much larger 
than $\Lambda_{\rm QCD}$.  Therefore, 
top quark production can be calculated to high precision using 
perturbative methods.  

Schematically, the $t\bar{t}$ cross section $\sigma_{t\bar{t}}$
can be expressed as a series: 
\begin{equation}
\sigma_{t\bar{t}}\sim{v}\sum_{k,i}\left(\frac{\alpha_s}{v}\right)^k(\alpha_s\ln{v})^i\left\{
\begin{array}{ll}
1&{\rm (LL)}\\
\alpha_s, v&{\rm (NLL)}\\
\alpha_s^2, \alpha_s{v}, v^2&{\rm (NNLL)}
\end{array}
\right..
\end{equation}
The velocity renormalization group (VRG)\cite{vrg} summation of 
$\alpha_s\ln{v}$ and Coulomb summation of $\alpha_s/v$ are performed to 
all orders.  
SM QCD calculations have been performed at NNLL 
order\cite{iain1}, and the remaining theoretical uncertainties 
are likely to be at the few percent level.  

Supersymmetry (SUSY) is the most prominent extension of the SM, and 
gives new contributions to top quark production through loop corrections.  
Supersymmetric one-loop corrections to the process 
$e^+e^-\rightarrow{t}\bar{t}$ above the 
threshold region have been studied in \cite{hollik}.
For scales below $m_t$, heavy super particles decouple
and they only affect $t\bar{t}$ pair production 
in the threshold region through matching conditions
at the scale $\mu=m_t$.  For heavy super particles,  
the supersymmetric contribution is  of the order of 
$\sigma_{t\bar{t}}\sim{v}\alpha_s(m_t/M_{\rm SUSY})^n$, 
where $M_{\rm SUSY}$ is the 
scale of the super particle masses, 
which is smaller than 1 TeV if supersymmetry plays a role in explaining
the hierarchy puzzle.  
Therefore, the leading SUSY correction 
could be comparable to the  SM NNLL corrections.   
In this paper, we will study the
supersymmetric contribution to top pair threshold production 
through one-loop matching 
conditions, neglecting subleading corrections which are suppressed by 
higher powers of $\alpha_s$ or $v$.  We do not consider SUSY loop corrections 
to the width of the top. 
They were considered in \cite{topdecay}.

For $t\bar{t}$ production in $e^+e^-$ annihilation via  
a photon or $Z$ boson exchange, the normalized 
cross section in the full QCD theory (at c.m. energy $\sqrt{s}$) is related to 
\begin{equation}
{\rm Im}\left[-i\int{d}^4{x}e^{iq\cdot{x}}
\langle{0}|T\mbox{\boldmath $J$}(x)\mbox{\boldmath $J$}^{\dagger}
(0)|0\rangle\right],
\label{eq:SMR}
\end{equation}
where $q=(\sqrt{s},0)$ and $\mbox{\boldmath $J$}$ 
is the production current.
In NRQCD, The non-relativistic $^3S_1$ current for top quark production 
near threshold can be written as \cite{iain1}
\begin{equation}
\mbox{\boldmath $J$}=c_1\mbox{\boldmath $O_{p}$}_{,1}+
\ldots,
\end{equation}
where
$\mbox{\boldmath $O_{p}$}_{,1}=
\psi^\dagger{\mbox{\boldmath $_p$}}\mbox{\boldmath $\sigma$}(i\sigma_2)
\chi^{\ast}{\mbox{\boldmath$_{-p}$}}$
and the ellipses denote terms suppressed by powers of $v$. 
The correlator of the currents $\mbox{\boldmath $J$}$ in Eq.~(\ref{eq:SMR}) 
is replaced by the correlator of 
$\mbox{\boldmath $O_{p}$}_{,1}$, 
which is proportional to the coordinate space Green function 
${G}(0,0)$, with its momentum space Fourier transformation
$\tilde{G}(\mbox{\boldmath $p$},\mbox{\boldmath  $p$}^\prime)$
being a solution of the Schr\"{o}dinger equation\cite{iain1}: 
\begin{equation}
\left[\frac{\mbox{\boldmath $p$}^2}{m}
-E-i\Gamma_t\right]
\tilde{G}(\mbox{\boldmath $p$},\mbox{\boldmath $p$}^\prime)+
\int{D}^n\mbox{\boldmath $q$}\;
\mu_S^{2\epsilon}
\tilde{V}(\mbox{\boldmath $p$},\mbox{\boldmath $q$}) 
\tilde{G}(\mbox{\boldmath $q$},
\mbox{\boldmath $p$}^\prime)
=(2\pi)^n\delta^{(n)}(\mbox{\boldmath $p$}-\mbox{\boldmath $p$}^\prime),
\label{eq:green}
\end{equation}
where $m$ is the heavy quark pole mass, $E\equiv(\sqrt{s}-2m)$, 
$n\equiv{3}-2\epsilon$ and 
${D}^n\mbox{\boldmath  $q$}\equiv{e}^{\epsilon\gamma_E}
(4\pi)^{-\epsilon}{d}^n\mbox{\boldmath $q$}/(2\pi)^n$.
$\tilde{V}(\mbox{\boldmath $p$},\mbox{\boldmath $q$})$ is the color singlet 
quark-antiquark potential in NRQCD, with the leading term 
being just the Coulomb potential. 
Therefore, the cross section for 
$t\bar{t}$ production near threshold is proportional to the product of 
the coefficient $c_{1}$ with the associated Green function. 

The main ingredients in the calculation of the top pair production 
cross section using NRQCD are the potential 
$\tilde{V}(\mbox{\boldmath $p$},\mbox{\boldmath $q$})$ 
for a quark-antiquark pair
and the Wilson coefficient $c_{1}$ of the production current, 
which are obtained via matching to the coefficients of the relevant operators 
calculated at the scale $\mu=m_t$, then running down to lower scales 
using the VRG\cite{vrg}. Recent SM calculations at NNLL order 
\cite{iainc1,iainp} are likely to have 
reduced the theoretical uncertainties to a few percent level.   
In what follows, we will consider  
the SUSY contribution to both the potential for a quark-antiquark pair
and the Wilson coefficient $c_1$ of the $^3S_1$ current, 
through one-loop matching conditions. 

The  potential interaction in NRQCD is written as 
\begin{equation}
{\cal{L}}=-V_{\alpha\beta\lambda\tau}(\mbox{\boldmath  $p$}, 
\mbox{\boldmath  $p$}^\prime)
{\psi^\dagger{\mbox{\boldmath $_{p^\prime}$}}}_{\alpha}
\psi{\mbox{\boldmath $_p$}_\beta}
\chi^\dagger{\mbox{\boldmath $_{-p^\prime}$}_\lambda}
\chi^{}{\mbox{\boldmath $_{-p}$}_\tau},
\end{equation}
where $\alpha, \beta, \lambda, \tau$ are the color and spin indices.  
$\psi{\mbox{\boldmath $_p$}}$ and 
$\chi{\mbox{\boldmath $_p$}}$ are in the ${\rm SU}(3)_c$ representation of 
$\mbox{\boldmath $3$}$ and 
$\mbox{\boldmath $\bar{3}$}$, which annihilate a quark and an antiquark, 
respectively. 
The leading term in $V$ is \cite{iainp}:
\begin{equation}
V=(T^A\otimes{\bar{T}}^A)
\frac{{V}_c^{(T)}}{\mbox{\boldmath $k$}^2},
\end{equation}
where $\mbox{\boldmath $k$}=\mbox{\boldmath $p$}^\prime-\mbox{\boldmath $p$}$.
At leading order, 
the SUSY contribution to the Coulomb term 
${V}_c^{(T)}$ is always zero by ${\rm SU}(3)_c$ gauge invariance.  
Integrating out heavy supersymmetric particles generates gauge invariant 
local operators
at the top quark scale.  In the transition to NRQCD at leading order 
in $\alpha_s$,  only the kinetic term for the top quark
generates the Coulomb potential. 
The subleading 
terms in the potential could get contributions from supersymmetric loops. 
However, they are suppressed by powers of  
$v$ relative to the leading Coulomb potential. 

For more 
general types of new physics that enter above the electroweak scale, and 
contribute to top quark pair production via perturbative 
matching at $\mu=m_t$,
the Coulomb potential is also protected due to 
the requirement of ${\rm SU}(3)_c$ gauge invariance. 
Therefore, the shape of the top pair production
cross section, which is related only to the Green function
in Eq.~(\ref{eq:green}),
will stay almost unchanged,
allowing a precise determination of the top quark mass based solely
on the SM calculations, with uncertainties that may be as small as 
100 MeV \cite{top}. 

\begin{figure}
\PSbox{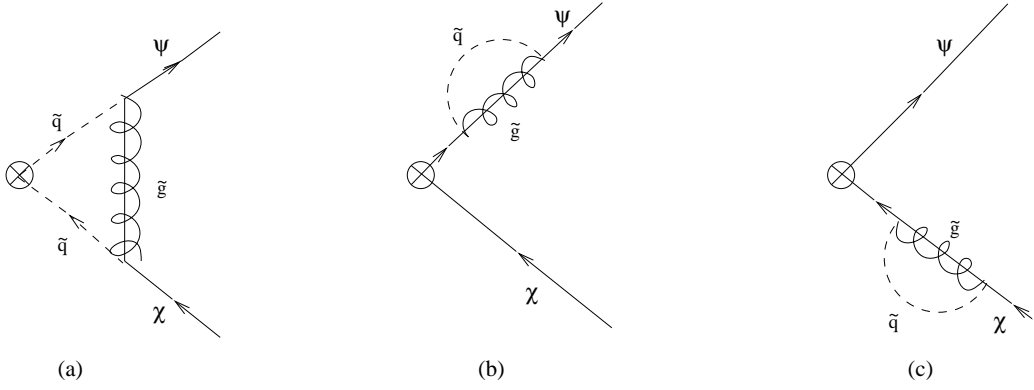 hoffset=40 voffset=20 hscale=80  
vscale=80}{6.8in}{2.2in}
\caption{One-loop supersymmetric correction of the $t\bar{t}$ 
pair production current.  The arrow shows the flow
of quark number and $\tilde{q}$ can be either $\tilde{t}_{1}$ or  
$\tilde{t}_{2}$.}
\label{fig:c1}
\end{figure}

The Feynman diagrams that contribute to the leading 
Wilson coefficient $c_1$ of the 
$^3S_1$ current are shown in Fig.~\ref{fig:c1}.
The SUSY corrections $\delta{c}_{1{\gamma}}$ or $\delta{c}_{1{Z}}$ 
to the production current associated with a photon or $Z$ boson 
are defined as:
\begin{eqnarray}
{\rm photon\;coupling:}\;\;\;&&
\mbox{\boldmath $A$}(\case{2}{3}e)\mbox{\boldmath $O_{p}$}_{,1}
(c_{1\gamma}+\delta{c}_{1{\gamma}});\\
Z{\rm \;coupling:}\;\;\;&&
\mbox{\boldmath $Z$}\sqrt{g_1^2+g_2^2}
({\case{1}{4}-\case{2}{3}\sin^2\theta_W})\mbox{\boldmath $O_{p}$}_{,1}
(c_{1Z}+\delta{c}_{1{Z}}),
\end{eqnarray}
where at leading order in perturbative QCD: $c_{1\gamma,Z}=1$.
We consider only the contributions from stop $\tilde{t}$ and gluino,
assuming that the flavor mixing between generations in the 
gluino-quark-squark couplings is small.  The 
contributions from ${\rm SU}(2)_L$, ${\rm U}(1)_Y$ gauginos
are suppressed by the electroweak couplings.  
Let $m_{\tilde{t}_{1,2}}$ denote the masses of the two physical 
stop mass eigenstates,
with the mixing angle $\theta$:
\begin{equation}
\left(
\begin{array}{l}
\tilde{t}_L\\\tilde{t}_R
\end{array} \right)=
\left(
\begin{array}{rl}
\cos\theta&\sin\theta\\
-\sin\theta&\cos\theta
\end{array} \right)\left(
\begin{array}{l}
\tilde{t}_1\\\tilde{t}_2
\end{array} \right),
\end{equation}
and $m_{\tilde{g}}$ denote the gluino mass.
The loop corrections to the production current associated with a photon or 
$Z$ boson
give (neglecting terms suppressed by 
higher powers of $v$) : 
\begin{eqnarray}
\delta{c}_{1{\gamma}}&=&C_F\frac{\alpha_s}{4\pi}
[F(m_{\tilde{t}_{1}},m_{\tilde{g}}, m_t) 
+F(m_{\tilde{t}_{2}},m_{\tilde{g}}, m_t) 
+G(m_{\tilde{t}_{1}},m_{\tilde{g}}, m_t)+
G(m_{\tilde{t}_{2}},m_{\tilde{g}}, m_t)]; \\
\delta{c}_{1{Z}}&=&C_F\frac{\alpha_s}{4\pi}\left\{\frac{1}
{\case{1}{4}-\case{2}{3}\sin^2\theta_W}\left[
(\case{1}{2}\cos^2\theta-\case{2}{3}\sin^2\theta_W)
F(m_{\tilde{t}_{1}},m_{\tilde{g}}, m_t)\right.\right. \nonumber \\
&+&\left.\left.(\case{1}{2}\sin^2\theta-\case{2}{3}\sin^2\theta_W)
F(m_{\tilde{t}_{2}},m_{\tilde{g}}, m_t) \right]
+G(m_{\tilde{t}_{1}},m_{\tilde{g}}, m_t)+
G(m_{\tilde{t}_{2}},m_{\tilde{g}}, m_t)\right\};
\end{eqnarray}
where
\begin{eqnarray}
F(m_{\tilde{t}_{i}},m_{\tilde{g}}, m_t)
&=&\int^1_0{\rm d}x\, 
\left[1-x+\frac{m^2_{\tilde{t}_{i}}+(x^2-1)m^2_t}
{m^2_{\tilde{t}_{i}}-m^2_{\tilde{g}}-m^2_t}\ln 
\frac{xm^2_{\tilde{t}_{i}}+(1-x)m^2_{\tilde{g}}+(x^2-x)m^2_t}
{m^2_{\tilde{t}_{i}}+(x^2-1)m^2_t}\right.\nonumber \\
&+&\left.x\ln \frac{xm^2_{\tilde{t}_{i}}+(1-x)m^2_{\tilde{g}}-x(1-x)m^2_t}
{(1-x)m^2_{\tilde{t}_{i}}+xm^2_{\tilde{g}}-x(1-x)m^2_t}\right];
\\
G(m_{\tilde{t}_{i}},m_{\tilde{g}}, m_t)&=&\int^1_0{\rm d}x\,
\frac{-2x(1-x)(xm^2_t\pm{m}_{\tilde{g}}m_t\sin2\theta)}
{xm^2_{\tilde{t}_{i}}+(1-x)m^2_{\tilde{g}}-x(1-x)m^2_t};
\;\;\;+\:{\rm for}\:i=1,\;\;-\:{\rm for}\:i=2;
\label{eq:fun_G}
\end{eqnarray}
and $C_F=4/3$ for ${\rm SU}(3)_c$.

In realistic SUSY models, 
for the squark mass matrix in the 
$(\tilde{q}_L,\tilde{q}_R)$ basis, the diagonal 
matrix elements $M_{LL,RR}^2\sim{M}^2_{\rm SUSY}$, 
while the off-diagonal matrix element
$M_{LR}^2\sim{M}_{\rm SUSY}m_t$.  
We chose the mass parameters randomly in the intervals: 
\begin{eqnarray}
(200\,{\rm GeV})^2\leq&{M}_{LL,RR}^2&\leq(1000\,{\rm GeV})^2, \nonumber \\
-(1000\,{\rm GeV})m_t\leq&{M}_{LR}^2&\leq(1000\,{\rm GeV})m_t, \nonumber \\
200\,{\rm GeV}\leq&|{m}_{\tilde{g}}|&\leq{1000}\,{\rm GeV},
\end{eqnarray}
with the requirement that the physical stop masses stay within the 
range of $200\,{\rm GeV}\leq{m}_{\tilde{t}_1}<{m}_{\tilde{t}_2}
\leq{1000}\,{\rm GeV}$.  
\begin{figure}
\PSbox{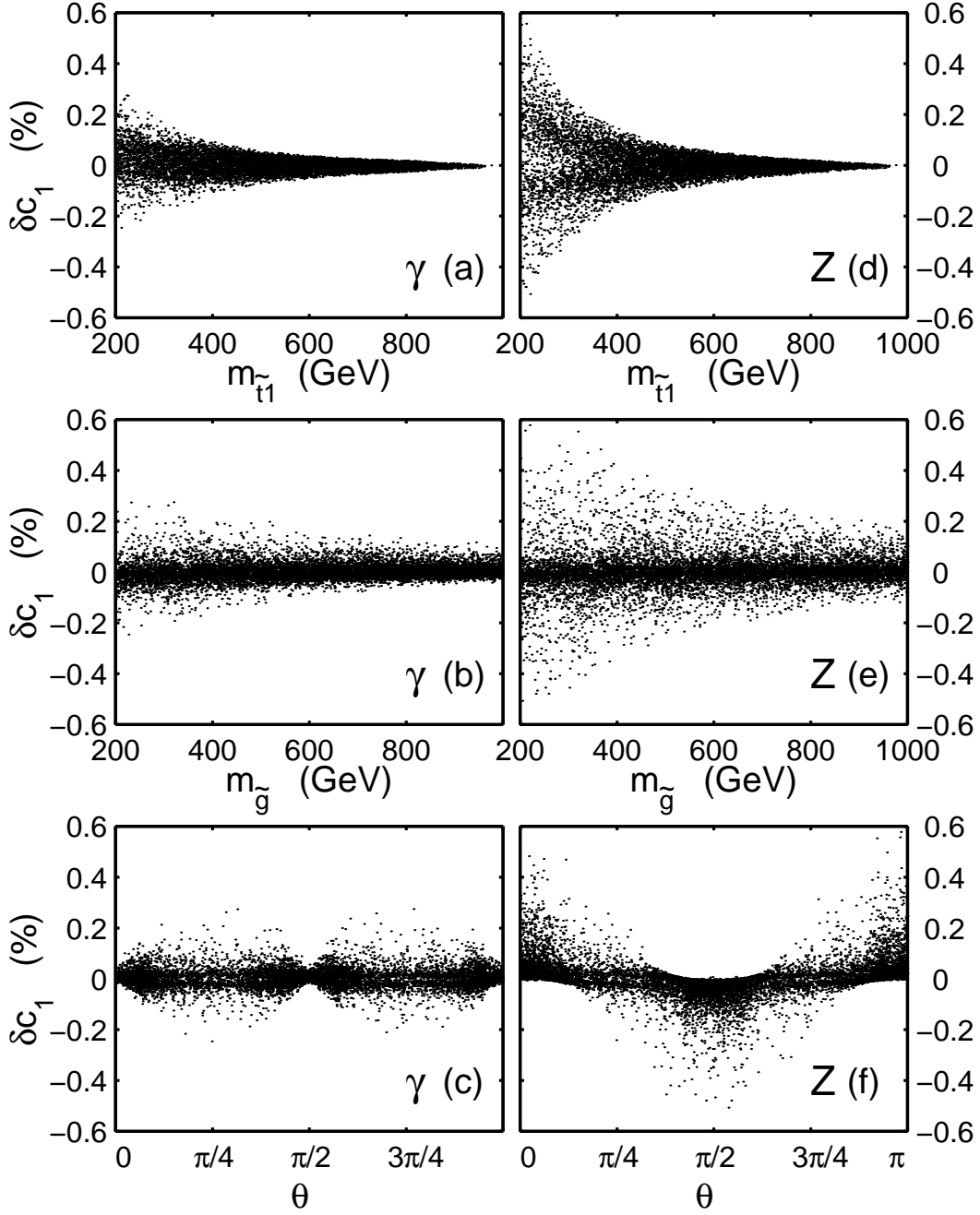 hoffset=-40 voffset=-220 hscale=120  
vscale=120}{8in}{6in}
\caption{Scatter plot of SUSY correction to 
$c_1$, for the currents associated with a 
photon [Fig.~(a), (b), (c)] and a $Z$ boson [Fig.~(d), (e), (f)].
$m_{\tilde{t}_1}$ and $m_{\tilde{g}}$ are the masses of the lighter 
stop and gluino, respectively, and $\theta$ is the stop mixing angle.  
The range of the SUSY mass parameters are chosen as 
described in the paper. }
\label{fig:scatter}
\end{figure}
The numerical results for the SUSY correction to the Wilson coefficient $c_1$
is shown in Fig.~\ref{fig:scatter} for the currents associated with a 
photon [(a), (b) and (c)] 
and a $Z$ boson [(d), (e) and (f)], respectively. 
The corrections relative to the tree level value  
are small in both cases: $-0.2\%<\delta{c}_{1\gamma}<0.3\%$ with 
97$\%$ of the points falling between $\pm{0.1}\%$ for the photon, and 
$-0.6\%<\delta{c}_{1Z}<0.6\%$ with 
86$\%$ of the points falling between $\pm{0.1}\%$ for the $Z$. 
The enhancement in the correction to the $Z$ current is due to the mixing 
between the $\tilde{t}_L$ and $\tilde{t}_R$, which alleviates the partial 
cancellation between the tree level left and right handed $Z$-current.

$\delta{c}_1$ goes to zero for large 
squark and gluino masses, since the SUSY contributions decouple 
for heavy super particles.  However, the gluino decouples slower, 
with $\delta{c}_1\sim{m}_t/|m_{\tilde{g}}|$ [unlike 
$\delta{c}_1\sim(m_t/m_{\tilde{q}})^2$ for the squark], as can be seen 
in $G(m_{\tilde{t}_{i}},m_{\tilde{g}}, m_t)$ in Eq.~(\ref{eq:fun_G}). 
The largest contribution comes from the term 
proportional to $\sin{2}\theta$.  For the $Z$ 
current, the contribution from the other 
terms are enhanced due to the 
left-right mixing in the stop sector, which causes the shift in the 
$\theta$ dependence in Fig.~\ref{fig:scatter}(f). 

The SUSY corrections are small compared with the 
${\cal{O}}(\alpha_s^2)$ NNLL 
SM contribution to $c_1$ \cite{c1NNLL}.
They are also smaller than the theoretical uncertainties 
in SM predictions at NNLL 
level \cite{iain1}, which are about $2-3\%$ due to the variation in 
$\nu$, for soft and ultrasoft renormalization scales 
$\mu_S=m_t\nu$ and $\mu_{US}=m_t\nu^2$.
Therefore, the corrections caused by the existence of heavy supersymmetric 
particles are negligible and it is sufficient to extract 
the top quark mass, decay width and top Yukawa 
coupling from the $t\bar{t}$ pair production 
near threshold based on SM calculations.

In summary, we calculated the leading SUSY contribution to top-antitop 
threshold production via its corrections to the matching condition for the 
potential and the Wilson coefficient $c_1$ of the 
$^3S_1$ production current associated with a photon or $Z$ boson  using the 
NRQCD framework.  The correction to the 
potential is zero (at leading order in $v$) due to ${\rm SU}(3)_c$ gauge 
invariance,
which is true in general for any new physics that
enters above the electroweak scale.  The shape of the top pair 
production cross section near threshold is unaffected, which can therefore
lead to a precise determination of the top quark mass 
based on the SM calculations.  The correction to the
coefficient $c_1$ of the production 
current is within $\pm{0.1}\%$ in most cases,
which is smaller than the SM NNLL corrections and the theoretical 
uncertainties from higher orders in QCD perturbation theory.
    
We have only considered the leading SUSY contribution from the gluino and stop,
neglecting terms proportional to higher powers of $\alpha_s$ and $v$. 
Flavor mixing between generations is usually small and not taken 
into consideration here.  
Neutralino, chargino and other squarks would contribute through electroweak 
loops, however their corrections are suppressed by the weak 
coupling constant instead of strong coupling.  
If there are super particles lighter than 
top quark, additional contributions to the running of the quark-antiquark 
potential and  the Wilson coefficient of the  
production current from $m_t$ to lower scales should be taken into account.

\acknowledgements
We would like to thank I. Stewart for useful discussion on $t\bar{t}$
threshold production.  S.S. and M.B.W. are supported by the DOE under grant 
DE-FG03-92-ER-40701.



\end{document}